\documentstyle[amssymb,psfig,epsfig,12pt]{article}

 0

\begin{document}

\begin{center}
{\large {Spin effects within the instanton model} \footnote{{\large Based on
the talks presented at the Advanced Study Institute "Symmetries and Spin"
(Praha-SPIN-2001), Prague, July 2001 and the workshop "Physical perspectives
of the JINR nuclotron", Varna, September 2001.}}\\[0.5cm]
} {\large {A.~E.~Dorokhov\\[0.5cm]
{\small {\it Bogoliubov Laboratory of Theoretical Physics, Joint Institute
for Nuclear Research, \\[0pt]
141980 Dubna, Moscow Region, Russia } \\[0.5cm]
}} }
\end{center}

\abstract{
Nonperturbative nolocal structure of QCD vacuum is well described by
instanton model. Specific helicity and flavor structure of zero modes of
quarks in instanton field allows simultaneously to explain some important
features of low- and high- energy hadron phenomemology. The basic
characteristics of hadron spectrum, partonic sum rules, heavy-quark
potential are briefly discussed within the instanton liquid model.}

\section{ \ \ \ \ \ Introduction}


Main features of {\it strong interaction} was defined long before the QCD
has been discovered. They are quarks and strangeness; symmetries and color;
spontaneous breaking of chiral symmetry; and, finally, confinement. QCD is
the non-abelian gauge theory of quarks and gluons called for solving these
problems. In the region of small distances (large energies) perturbative
theory of QCD (pQCD) works well with very high precision and describes the
scaling and its logarithmic violation. In particular, pQCD describes the
evolution of parton distributions (extracted from experimental data and
rescaled to one scale) with $Q^{2}$ at high $Q^{2}$. Thus, the hadron
processes at (asymptotically) high energies are described by pQCD. In the
very low $Q^{2}$ region they are fixed by symmetries of strong interaction
(low energy theorems, chiral quark models, etc.). Still the main phenomena
of strong interaction are not explicitly explained. There are problems with
description of hadron spectrum, hadron distribution functions of quarks and
gluons at intermediate energies, {\it etc.} The transition region between
very low and high energies is very interesting aspect of modern experimental
and theoretical searches. It is this region where different nonperturbative
approaches was suggested: Lattice QCD, QCD sum rules, effective field
models, etc. We are going to consider the instanton model of QCD vacuum and
some of its applications.

\section{The instanton model of QCD vacuum}

To describe the strong interactions not only at short distances but also at
medium and long distances one needs to understand how to calculate the
fundamental elements of the theory, its Green functions. This task may be
done within the lattice QCD simulations \cite{Bec00b}-\cite{CuZwan00} or
effective QCD models like the instanton model of the QCD vacuum \cite
{Shuryak96}. Lorentz invariance allows us to decompose the full quark
propagator into Dirac vector and scalar pieces
\begin{equation}
S^{-1}(p^{2})=Z^{-1}(p^{2})[i\gamma \cdot p+M(p^{2})].  \label{Sq1}
\end{equation}
Asymptotic freedom means that, as $p^{2}\rightarrow \infty $, $%
S^{-1}(p^{2})\rightarrow i\gamma \cdot p+m,$ (the free propagator) where $m$
is the bare quark mass. The gluon propagator in the Landau gauge is given by
\begin{equation}
D_{\mu \nu }^{ab}\left( p^{2}\right) =\delta ^{ab}\left( \delta ^{\mu \nu }-%
\frac{p_{\mu }p_{\nu }}{p^{2}}\right) D\left( p^{2}\right)  \label{Dg}
\end{equation}
with asymptotic large $p^{2}$ behavior $D\left( p^{2}\right) \rightarrow
p^{-2}$. The explicit form of nonperturbative functions $M(p^{2}),$ $D\left(
p^{2}\right) $ etc. may be found within the instanton vacuum model.

The vacuum gauge field is taken to be the sum of individual instanton fields
with their centra at positions $z_{j}$. Each instanton is in the singular
gauge
\begin{equation}
A_{\mu }^{a}(x)\frac{\tau ^{a}}{2}=\frac{1}{g}\tau ^{a}\eta _{\mu \nu
}^{a}(x-z_{j})_{\nu }\phi (x-z_{j}),  \label{Inst_Field}
\end{equation}
where $\eta _{\mu \nu }^{a}$ is the 't Hooft symbols and $\phi (x)$ is the
profile function.

The instanton vacuum fluctuations are responsible for the nonperturbative
effects at intermediate scale of order 0.3 fm. The longer distances are
described by the inter-instanton interaction and noninstantonic
nonperturbative fluctuations that generate confinement of gluons and quarks.
All these effects, by assumption, lead to stabilization of the instanton
ensamble and the average density of instantons $n\approx 1$ fm$^{4}$ with
the instanton radii $\rho \approx 0.3$ fm \cite{Shuryak96}. Another effect
of the confining background is the modification of the shape of instanton
solution at large distances in the form of the constrained instanton
solution \cite{Espin90,DEMM99}. The ansatz for the profile function for the
constrained instanton field \cite{DEMM99} in the singular gauge may be
chosen as
\begin{equation}
\phi _{I}(x)=\frac{\overline{\rho }^{2}(x^{2})}{x^{2}\left( x^{2}+\overline{%
\rho }^{2}(x^{2})\right) }.  \label{Inst_Profile}
\end{equation}
The standard instanton solution is obtained with the constant size of
instanton $\overline{\rho }^{2}(x^{2})=\rho ^{2}$. For the constrained
instantons one uses exponentially-decreasing functions $\overline{\rho }%
^{2}(x),$ normalized as $\overline{\rho }^{2}(0)=\rho ^{2}$. The constrained
quark zero mode \cite{Espin90} is given by
\begin{equation}
\psi ^{\pm }(x)=\sqrt{2}\varphi (x)\frac{\widehat{x}}{\left| x\right| }\chi
^{\pm },\qquad \varphi (x)=\frac{\overline{\rho }(x^{2})}{\pi (x^{2}+%
\overline{\rho }^{2}(x^{2}))^{3/2}},  \label{PsiReg}
\end{equation}
where $\chi $ is a color Dirac spinor given by $\chi ^{\pm }\overline{\chi }%
^{\pm }=\left( \gamma _{\mu }\gamma _{\nu }/16\right) \left( 1\pm \gamma
_{5}\right) /2\tau _{\mu }^{\pm }\tau _{\nu }^{\mp }$ and $\tau _{\nu }^{\pm
}=\left( \mp i,\overrightarrow{\tau }\right) $, with the upper (lower) signs
corresponding to solutions in the instanton (anti-instanton) field. We shell
use the form \cite{DEMM99}
\begin{equation}
\overline{\rho }^{2}(x^{2})=\rho ^{2}\frac{2}{\Gamma (1/3)3^{1/3}}\left(
\frac{x}{R}\right) ^{2}K_{4/3}\left( \frac{2}{3}\left( \frac{x^{2}}{R^{2}}%
\right) ^{3/4}\right) ,  \label{rhoX}
\end{equation}
where $K_{\nu }\left( z\right) $ is the modified Bessel function. The
specific feature of the constrained instanton is that at small distances it
is close to the standard instanton profile of size $\rho $, and at large
distances it has exponentially-decreasing asymptotics governed by a
large-scale parameter $R$, such as $\rho <R$. These shapes are motivated by
considering modifications of the instanton in the background field of
large-scale vacuum fluctuations \cite{DEMM99}. The constrained instanton
profile, as opposed to the unconstrained one, provides the correct
large-distance asymptotics of the quark and gluon field correlators \cite
{DEMM99}.

The dynamical quark mass generated by the instanton background is expressed
through the 4-dimensional Fourier transform of the quark zero mode profile $%
\widetilde{\varphi }(p)$ \cite{CDG78}
\begin{equation}
M(p^{2})=Cp^{2}\widetilde{\varphi }^{2}(p^{2}),  \label{Mp2}
\end{equation}
where the constant $C>0$ is determined from the gap equation \cite
{CDG78,DP86}
\begin{equation}
\int \frac{d^{4}k}{(2\pi )^{4}}\frac{M^{2}(k)}{k^{2}+M^{2}(k)}=\frac{n_{c}}{%
4N_{c}},  \label{Gap2}
\end{equation}
with the instanton density $n_{c}=0.0016$ GeV$^{4}$ \cite{Shuryak96}.

The single instanton contribution to the gluon propagator
\begin{eqnarray}
G_{\mu \nu }^{CI,ab}(p) &\equiv &\int d^{4}xe^{ipx}\left\langle 0\left|
A_{\mu }^{a,I}(x)A_{\nu }^{b,I}(0)\right| 0\right\rangle _{I}=\delta
^{ab}\left( \delta ^{\mu \nu }-\frac{p^{\mu }p^{\nu }}{p^{2}}\right)
G^{CI}(p),  \nonumber \\
G^{CI}(p) &=&-\frac{n_{c}}{N_{c}^{2}-1}p^{2}\widetilde{\phi }^{2}(p^{2}),
\label{GProp_I}
\end{eqnarray}
is expressed through the Fourier transform of the instanton profile function
$\widetilde{\phi }(p^{2})$. Summing the contribution to the propagator of
any number of instantons, which is analog of a self-energy resummation in
perturbative theory, we get the dressed in the instanton vacuum gluon
propagator in the Landau gauge
\begin{equation}
G_{\mu \nu }^{ab}(p)=\frac{\delta ^{ab}}{p^{2}+M_{CI}^{2}(p^{2})}\left(
\delta ^{\mu \nu }-\frac{p^{\mu }p^{\nu }}{p^{2}}\right) ,
\label{GProp_Dressed}
\end{equation}
with a dynamical gluon mass
\[
M_{CI}^{2}(p^{2})=p^{4}G^{CI}(p).
\]
It is important that the Green function of the effective model have correct
large $p^{2}$ behavior and models nonperturbative dynamics at large
distances. Also, the result of instanton model (\ref{Mp2}) and (\ref
{GProp_Dressed}) are close to the results of lattice calculations \cite
{Bec00b}-\cite{CuZwan00} in the infrared region.

In general instanton induced interaction leads to effective fermion
interaction which is for two flavors becomes
\[
L_{2}=\frac{2N_{c}^{2}}{n}\int \frac{d^{4}k_{1}d^{4}k_{2}d^{4}l_{1}d^{4}l_{2}%
}{(2\pi )^{12}}\sqrt{M(k_{1})M(k_{2})M(l_{1})M(l_{2})}
\]
\[
\cdot \frac{\epsilon ^{f_{1}f_{2}}\epsilon _{g_{1}g_{2}}}{2(N_{c}^{2}-1)}%
\left[ \frac{2N_{c}-1}{2N_{c}}(\psi _{Lf_{1}}^{\dagger }(k_{1})\psi
_{L}^{g_{1}}(l_{1}))(\psi _{Lf_{2}}^{\dagger }(k_{2})\psi
_{L}^{g_{2}}(l_{2}))\right.
\]
\begin{equation}
\left. +\frac{1}{8N_{c}}(\psi _{Lf_{1}}^{\dagger }(k_{1})\sigma _{\mu \nu
}\psi _{L}^{g_{1}}(l_{1}))(\psi _{Lf_{2}}^{\dagger }(k_{2})\sigma _{\mu \nu
}\psi _{L}^{g_{2}}(l_{2}))+\left( L\rightarrow R\right) \right] .
\label{tHooft}
\end{equation}
This is so called effective 't Hooft vertex.

\section{Instanton forces and spectroscopy.}

The QCD vacuum has quite complicated structure. Conventionally the
nonperturbative fields can be divided by two parts: short wave component
which provides the interaction of quarks at small distances and long wave
one which respects for the confinement. In the framework of instanton liquid
model the first part is connected with short distance vacuum correlations
where the single-instanton contribution with effective size $\rho
_{c}\propto 1.5\div 2\ GeV^{-1}$ dominates. Second component is related with
long wave collective excitations of instanton liquid and noninstantonic
vacuum fluctuations with wave length $R\approx R_{conf}$, where $R\approx
3\rho _{c}$ is average distance between instantons and $R_{conf}\approx
5\div 6\ GeV^{-1}$ is a confinement radius.

The hadron model based on these assumptions was considered in \cite
{DK92,ShRos89}. Within this model the hadron energy is
\begin{equation}
E_{H}=\frac{1}{R}\sum_{i={\rm flavor}}N_{i}\omega \left( m_{i}R\right) +E_{%
{\rm vac}},  \label{E_H}
\end{equation}
where the first term is a sum of kinetic energy of quarks confined in the
hadron bag with energy $\omega \left( m_{i}R\right) /R$ and the second term
takes into account the quark interaction with physical vacuum. The
interaction with long-ranged vacuum fluctuations (condensates) has form of
power corrections
\begin{equation}
E_{{\rm cond}}=-R^{2}\sum_{i={\rm flavor}}N_{i}C^{QQ}\left( m_{i}R\right)
\left\langle 0\left| \overline{q}_{i}q_{i}\right| 0\right\rangle +\sim
\left\langle 0\left| G_{\mu \nu }^{a}G_{\mu \nu }^{a}\right| 0\right\rangle
R^{3}+...  \label{E_cond}
\end{equation}
and stabilizes the bag balancing the internal (kinetic energy) and external
(vacuum energy) pressures $\partial E_{H}/\partial R=0.$ By using the
stabilization condition the hadron mass scale is defined by the quark
condensate as
\begin{equation}
E_{H}=\frac{3}{2}N_{q}\left( \frac{2\pi \omega ^{2}}{24\left( \omega
-1\right) }\right) ^{1/3}\left| \left\langle 0\left| \overline{q}q\right|
0\right\rangle \right| ^{1/3}\approx \left\{
\begin{array}{c}
750\quad {\rm MeV\qquad for\quad mesons} \\
1100\quad {\rm MeV\qquad for\quad baryons}
\end{array}
\right. .  \label{Escale}
\end{equation}
This estimate for the mass scale of the nonstrange hadrons is consistent
with QCD sum rule estimates \cite{BI82}.

Due to specific structure of the effective 't Hooft interaction (\ref{tHooft}%
), the interaction of quarks through the exchange by small size instantons
given by
\begin{equation}
\Delta E_{{\rm inst}}=\frac{\beta \rho _{c}}{R^{3}}\sum_{i\neq j}\frac{%
N_{ij}I\left( m_{i,}m_{j}\right) }{m_{i}^{\ast }m_{j}^{\ast }}\left( 1-%
\overrightarrow{\sigma }_{i}\overrightarrow{\sigma }_{j}\right)
\label{dEinst}
\end{equation}
produces spin-spin forces in the hadron multiplets and solves the $%
U_{A}\left( 1\right) $ problem
\[
\Delta E_{{\rm inst}}^{{\rm p}}=-\frac{3}{2}\frac{\lambda _{0}}{R^{3}}%
,\qquad \Delta E_{{\rm inst}}^{{\rm \pi }}=-\frac{\lambda _{0}}{R^{3}}%
,\qquad \Delta E_{{\rm inst}}^{{\rm \eta }}=\frac{\lambda _{0}-4\lambda _{s}%
}{3R^{3}},
\]
\[
\Delta E_{{\rm inst}}^{{\rm \Delta }}=0,\qquad \Delta E_{{\rm inst}}^{{\rm %
\rho }}=0,\qquad \Delta E_{{\rm inst}}^{{\rm \eta }^{\prime }}=\frac{\lambda
_{0}+2\lambda _{s}}{3R^{3}}.
\]
By using standard vacuum parameters $\left\langle 0\left| \overline{q}%
_{i}q_{i}\right| 0\right\rangle \approx -\left( 250\quad {\rm MeV}\right)
^{3}$ and $\rho _{c}\approx 2$ {\rm GeV}$^{-1}$ the satisfactory results for
ground state hadrons have been obtained \cite{DK92}.

\section{The axial anomaly, the nucleon structure and the parton sum rules}

The deep inelastic lepton - nucleon scattering processes (DIS) occurring at
small distances characterize the internal structure of the elementary
particles. In the past decade new experimental data with high precision and
in large kinematic region has become available.

This is primarily a result of the SLAC - EMC - SMC - HERMES measurements
\cite{SLAC1,EMC,SMC,SLAC2,SMC98} of helicities of the charged constituents
of the proton and neutron. The EMC data analysis\cite{EMC} has resulted in a
striking conclusion: the sum of the helicities of the quarks inside a
proton, $\Delta \Sigma $, was found to be extremely small, $(\Delta \Sigma
<<1)$, and the Ellis-Jaffe (EJ) {\cite{EJSR}} sum rule (SR) is strongly
violated.

Then, it was concluded from NMC data analysis of the unpolarized structure
function of nucleon $F_{2}(x,Q^{2})$ \cite{NMC,E866} that $u-$ quark sea in
the proton is suppressed w.r.t. $d-$ quark sea, that is the Gottfried (G) SR
\cite{GtSR} is violated, too. All this is in dramatical contradiction with
the expectation of the naive parton model where all these sum rules are
fulfilled. From the other side, the polarization experiments on neutron -
contained targets confirmed that fundamental Bjorken (Bj) SR \cite{BjSR} is
valid \cite{SLAC2,SMC98}.

In this part we want to argue \cite{DKSpin} that the observed inconsistency
in parton sum rules are a manifestation of nonperturbative structure of the
QCD vacuum. Within the framework of this approach the breaking mechanism of
QCD partonic SR is connected with a mixture of sea quarks with large
transverse momentum in the nucleon wave function. This quark sea results
from scattering valence quark off nonperturbative vacuum fluctuation,
instanton.

This interaction in the limit of small size instanton is defined by the
effective 't Hooft Lagrangian that may be written in the form:
\begin{equation}
{\cal L}^{inst}(x)=(2n_{c}k^{N_{f}})\Re e\det (q_{R}q_{L})  \label{e2}
\end{equation}
with the anomaly equation given by $\partial _{\mu }j_{\mu
}^{5}=-2N_{f}(2n_{c}k^{N_{f}})\Im m\det (q_{R}q_{L}),$ where $k=\frac{%
\displaystyle4\pi \rho _{C}^{3}}{\displaystyle3}\frac{\displaystyle\pi }{%
\displaystyle(m_{\ast }\rho _{C})}$ is the effective instanton - quark
coupling and  $m_{\ast }=m-2/3\rho _{c}^{2}<~0~|\bar{Q}Q|0>$ is the
effective quark mass in the physical vacuum.

It is important that the instanton induced interaction (\ref{e2}) changes
the chirality of a quark by the value $\Delta Q=-2N_{f}$ and acts only for
differently flavored quarks. From this it immediately follows that sea
quarks have negative helicity and screen the helicity of a valence quark on
which they are produced. On instanton the sea pair in the state with Right
chirality is created and on the anti-instanton the quark pair with Left
chirality is appeared.

Another thing is that on u-(d-) quark only $d\bar{d}-(u\bar{u}-)$ and $s\bar{%
s}-$ quark sea is possible. Therefore there is more{\bf \ }$d${\bf $-$ }sea
quarks in the proton. As a result it turns out that in the framework of the
instanton mechanism the spin and flavor structure of nucleon quark sea is
strongly correlated with the spin-flavor of the valence nucleon wave
function.

Thus, specific helicity and flavor structure of quark zero mode interaction
in the instanton field allows us simultaneously to explain the breaking of
both Ellis-Jaffe SR related to significant breaking of quark helicity
conservation and the Gottfried SR caused by the violation of the $SU_{f}(2)$%
- symmetry of quark sea.

I should note that a perturbative quark-gluon vertex does not flip the
helicity neither does feel the flavor of the valence quark. Thus, within the
perturbative QCD it is not possible, in principle, to explain the
experimentally observed significant violation of both sum rules.

From the vertex (\ref{e2}) we obtain the anomalous instanton contributions
to different PSR and axial charges \cite{DAdel}:

{\it Gottfried sum rule}
\[
\Delta S_{G}=-\frac{2}{3}\left( \overline{d}-\overline{u}\right) =-\frac{2}{3%
}a;
\]
flavor triplet and octet axial constants
\[
\Delta g_{A}^{3}=\Delta u-\Delta d=-\frac{10}{3}a_{s};
\]
\[
\Delta g_{A}^{8}=\Delta u+\Delta d-2\Delta s=-4a+2a_{s};
\]
flavor singlet axial charge
\begin{equation}
\Delta \Sigma _{inst}\equiv \Delta g_{A}^{0}=\Delta u+\Delta d+\Delta
s=-4a-4a_{s};  \label{Dsq}
\end{equation}
Ellis-Jaffe integrals for proton and neutron
\[
\Delta S_{EJ}^{p}=\sum_{p}e_{q}^{2}\Delta q/2=-\frac{1}{9}(5a+6a_{s});
\]
\[
\Delta S_{EJ}^{n}=\sum_{n}e_{q}^{2}\Delta q/2=-\frac{1}{9}(5a+a_{s});
\]
Bjorken integral
\[
\Delta S_{Bj}=\Delta S_{EJ}^{p}-\Delta S_{EJ}^{n}=-\frac{5}{9}a_{s}=\frac{1}{%
6}\Delta g_{A}^{3},
\]
where $a$ ($a_{s}\lesssim a/2$) is the probability to create nonstrange
(strange) sea quark pair in instanton field. If we attribute all $SU_{f}(2)$
asymmetry of sea measured by the E866 Collab. \cite{E866}
\begin{equation}
\int\limits_{0}^{1}dx\ [\bar{d}(x)-\bar{u}(x)]=0.118\pm 0.012,  \label{NMC}
\end{equation}
to the instanton contribution then we obtain the value for the coupling $%
a=0.118\pm 0.012$. The parameter $a_{s}$ measures the difference $%
g_{A}^{8}-\Delta \Sigma =-3\Delta s$ and is given in our model by $6a_{s}$.
From experiment it is estimated \cite{Ji01spin} $\Delta s=-0.14\pm 0.03$ and
thus $a_{s}=0.07\pm 0.015$.

To estimate the (nonanomalous) valence quark contribution we shall use the
quark model where relativistic effect reduces the helicity of quarks with
respect to the nonrelativistic quark model result and take as a conservative
estimation the value
\begin{equation}
\Delta u_{v}+\Delta d_{v}=0.8\pm 0.15  \label{Dvq}
\end{equation}
From (\ref{Dsq}), (\ref{NMC}), (\ref{Dvq}) we obtain the final result for
the singlet axial charge of the proton:
\[
\Delta \Sigma =0.09\pm 0.25.
\]
Our theoretical error is the sum of indefinetness in experimental number for
GSR for sea quarks and relativistic effects for valence quarks. Physically
this compensation for the helicity of initial quark means a transformation
of the valence quark spin momentum into the angular momentum of quark pair
(in $O^{++}$ state) created by instanton. This very small value for the
total quark contribution to the proton's spin is in agreement with the
result of the latest analysis of the first moment of $g_{1}(x)$ structure
function \cite{Ji01spin}
\[
\Delta \Sigma =0.16\pm 0.08.
\]

It should be stressed that inspite of the fact that the instanton induced
interaction contributes to $g_{A}^{3}$ it does not violate the Bjorken sum
rule and violation of GSR is strongly correlated with contribution of axial
anomaly to EJSR.

Another predictions of the instanton model for parton distribution integrals
are:

1) Asymmetries in flavor structure of quark sea
\begin{equation}
\overline{d}-\overline{s}=2\left( 1-\frac{3}{2}\frac{a_{s}}{a}\right)
a=0.026\pm 0.069;\qquad \overline{u}-\overline{s}=2\left( 1-3\frac{a_{s}}{a}%
\right) a=-0.184\pm 0.11;  \label{SeaAsym}
\end{equation}

2) New sum rule that is saturated by nonperturbative asymmetries in the
nucleon sea
\begin{equation}
\frac{9}{25}\int_{0}^{1}dx\left[ g_{1}^{p}\left( x\right) -6g_{1}^{n}\left(
x\right) \right] +\frac{3}{2}\int_{0}^{1}\frac{dx}{x}\left[ F_{2}^{\mu
p}\left( x\right) -F_{2}^{\mu n}\left( x\right) \right] =\frac{3}{5}
\label{IntAsym}
\end{equation}
By using the E866 data for the GSR integral and the theoretical prediction
for the BjSR integral \cite{SMC98}
\[
\int_{0}^{1}dx\left[ g_{1}^{p}\left( x\right) -g_{1}^{n}\left( x\right) %
\right] =0.181\pm 0.003
\]
the model predict the neutron integral of $g_{1}^{n}\left( x\right) $ as
\[
\int_{0}^{1}dxg_{1}^{n}\left( x\right) =-0.085\pm 0.007,
\]
that may be compared with world average experimental value \cite{SMC98}
\[
\int_{0}^{1}dxg_{1}^{n}\left( x\right) =-0.075\pm 0.031.
\]

3) Sea quark spin asymmetries
\begin{equation}
\Delta \overline{u}-\Delta \overline{d}=\frac{5}{3}a=0.20\pm 0.02>0;\qquad
\Delta \overline{s}=-a_{s}=-0.07\pm 0.015<0.  \label{SeaSpinAsym}
\end{equation}
Thus we demonstrate that the instanton motivated nonperturbative sea
fluctuations of nucleon sea describe well the current experimental situation%
\footnote{%
In the papers \cite{DKSpin,DAdel} the model of the sea quark distributions
induced by instanton interaction has been considered.}. \

\section{The Instanton Effects on the Heavy Quark Potential.}

The gauge invariant potential between a very heavy quark and antiquark in a
color-singlet state is given by\cite{CDG78,CDG78b}
\begin{equation}
V\left( R\right) =-\lim_{T\rightarrow \infty }\ln \left\langle 0\left|
W\left( T,R\right) \right| 0\right\rangle ,  \label{V(R)}
\end{equation}
where $W\left( T,R\right) $ is a Wilson-loop $Tr\left( P\exp
ig\oint_{C\left( R,T\right) }dz^{\mu }A_{\mu }\left( z\right) \right) $,
with loop $C\left( R,T\right) $ being the rectangular closed curve with a
spatial length $R$ and a temporal length $T$. The origin of this expression
is in the fact that during scattering heavy quark off the instanton its wave
function changes as
\[
\Psi _{Q}^{\prime }\left( x\right) =\left[ O\left( \overrightarrow{x},%
\overrightarrow{p},\overrightarrow{\sigma }\right) U^{-1}\left(
\overrightarrow{x}\right) \right] \Psi _{Q}\left( x\right)
\]
with phase factor
\[
U^{-1}\left( \overrightarrow{x}\right) =P\exp \left[ ig\int_{-\infty
}^{\infty }dx_{4}A_{4}\left( \overrightarrow{x},x_{4}\right) \right]
\]
and spin-orbit operator
\[
O=1-\frac{i}{m_{Q}}\overrightarrow{L}\cdot \overrightarrow{\nabla }-\frac{i}{%
2m_{Q}}\overrightarrow{\sigma }\cdot \overrightarrow{\nabla }-\frac{1}{%
2m_{Q}^{2}}\left( \overrightarrow{L}\cdot \overrightarrow{\nabla }\right)
\left( \overrightarrow{\sigma }\cdot \overrightarrow{\nabla }\right) +\frac{1%
}{4m_{Q}^{2}}\overrightarrow{\sigma }\left[ \overrightarrow{p}\cdot
\overrightarrow{\nabla }\right] .
\]
Within the single-instanton approximation the phase factor may be reduced to
\[
U^{-1}\left( \overrightarrow{x}\right) =\cos \alpha \left( \overrightarrow{x}%
\right) -i\tau ^{a}n^{a}\sin \alpha \left( \overrightarrow{x}\right) ,
\]
where
\[
\alpha \left( \overrightarrow{x}\right) =\int_{-\infty }^{\infty }dx_{4}\phi
\left( \overrightarrow{x},x_{4}\right) ,\qquad n^{a}=\widehat{%
\overrightarrow{x}}.
\]
Then the quark-quark scattering potential is given by potential
\[
H_{QQ}=\left[ -1+O_{s-s}+O_{s-L}\right] V\left( \overrightarrow{x}_{1}-%
\overrightarrow{x}_{2}\right)
\]
with $V\left( R\right) =-2\int dn\left( \rho \right) \rho ^{3}W\left(
R\right) $ and expansion in scalar part
\[
W\left( R\right) =\frac{1}{N_{c}}\int d\overrightarrow{z}tr\left[ 1-U\left(
\frac{\overrightarrow{R}}{2}-\overrightarrow{z}\right) U^{-1}\left( \frac{%
\overrightarrow{R}}{2}+\overrightarrow{z}\right) \right] ,
\]
spin-spin part proportional to $\overrightarrow{\sigma }\cdot
\overrightarrow{\sigma }$%
\[
W_{s-s}\left( R\right) =\frac{1}{N_{c}}\left[ 2\frac{W^{\prime }\left(
R\right) }{R}+W^{\prime \prime }\left( R\right) \right]
\]
and tensor part proportional to $\left[ \overrightarrow{\sigma }_{i}\cdot
\overrightarrow{\sigma }_{j}-\frac{\delta _{ij}}{3}\overrightarrow{\sigma }%
_{i}\times \overrightarrow{\sigma }_{j}\right] \overrightarrow{x}_{i}%
\overrightarrow{x}_{j}$%
\[
W_{T}\left( R\right) =\frac{1}{N_{c}}\left[ \frac{W^{\prime }\left( R\right)
}{R}-W^{\prime \prime }\left( R\right) \right] .
\]

The instanton and constrained instanton model predictions are given in Figs
2-4. 
\begin{figure}[tbp]
\hspace*{1.cm}
\begin{minipage}{2.5in}
\vspace*{0.5cm}
\epsfxsize=6cm
\epsfysize=9cm
\centerline{\epsfbox{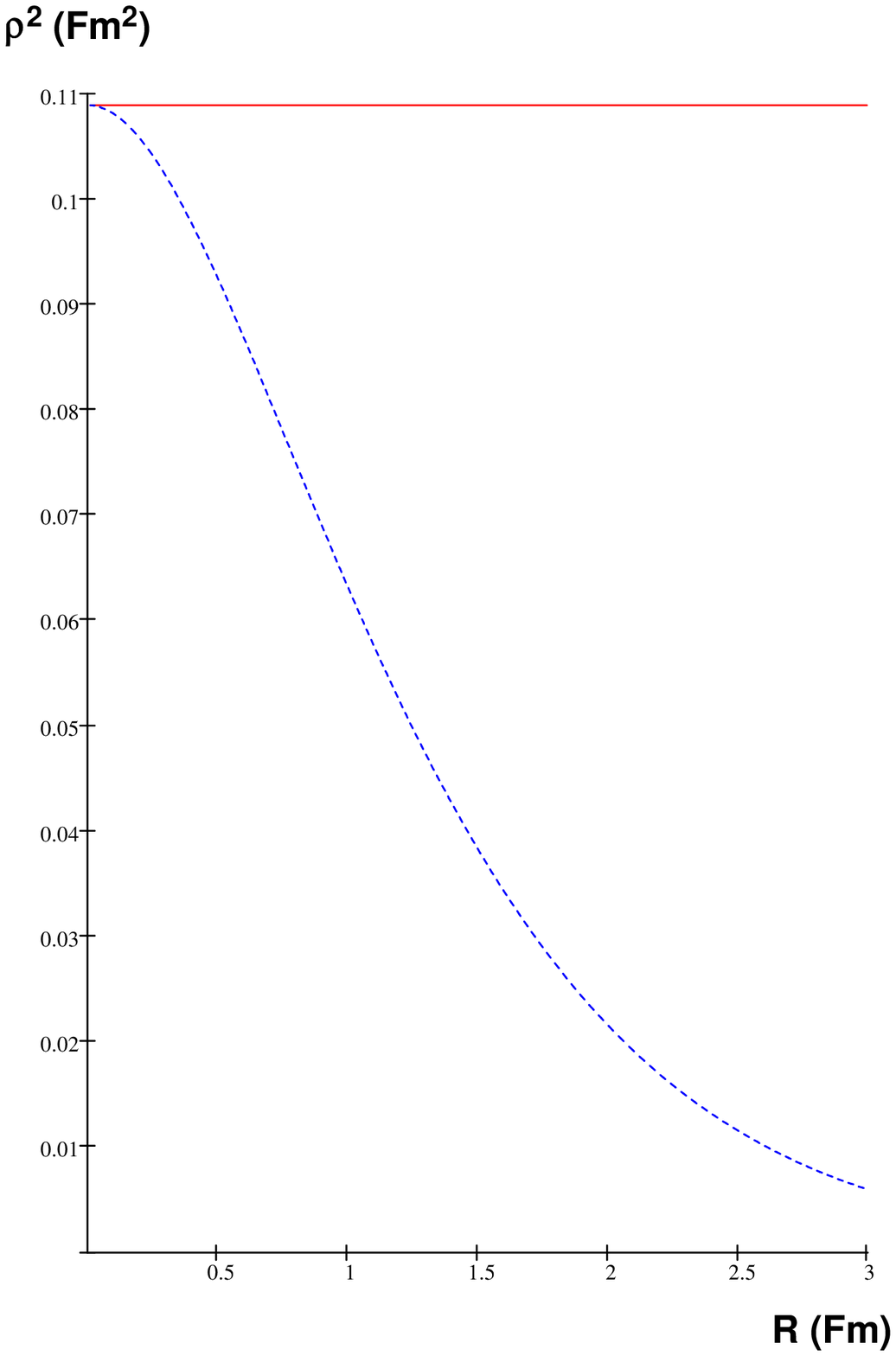}}
 \caption[dummy0]{
Instatnon size distribution.
The solid line is for instanton and
the dashed one is for constrained instanton.
  \label{fig:R2} }
\end{minipage}
\hspace*{0.5cm}
\begin{minipage}{2.5in}
\vspace*{0.5cm}
\epsfxsize=6cm
\epsfysize=9cm
\centerline{\epsfbox{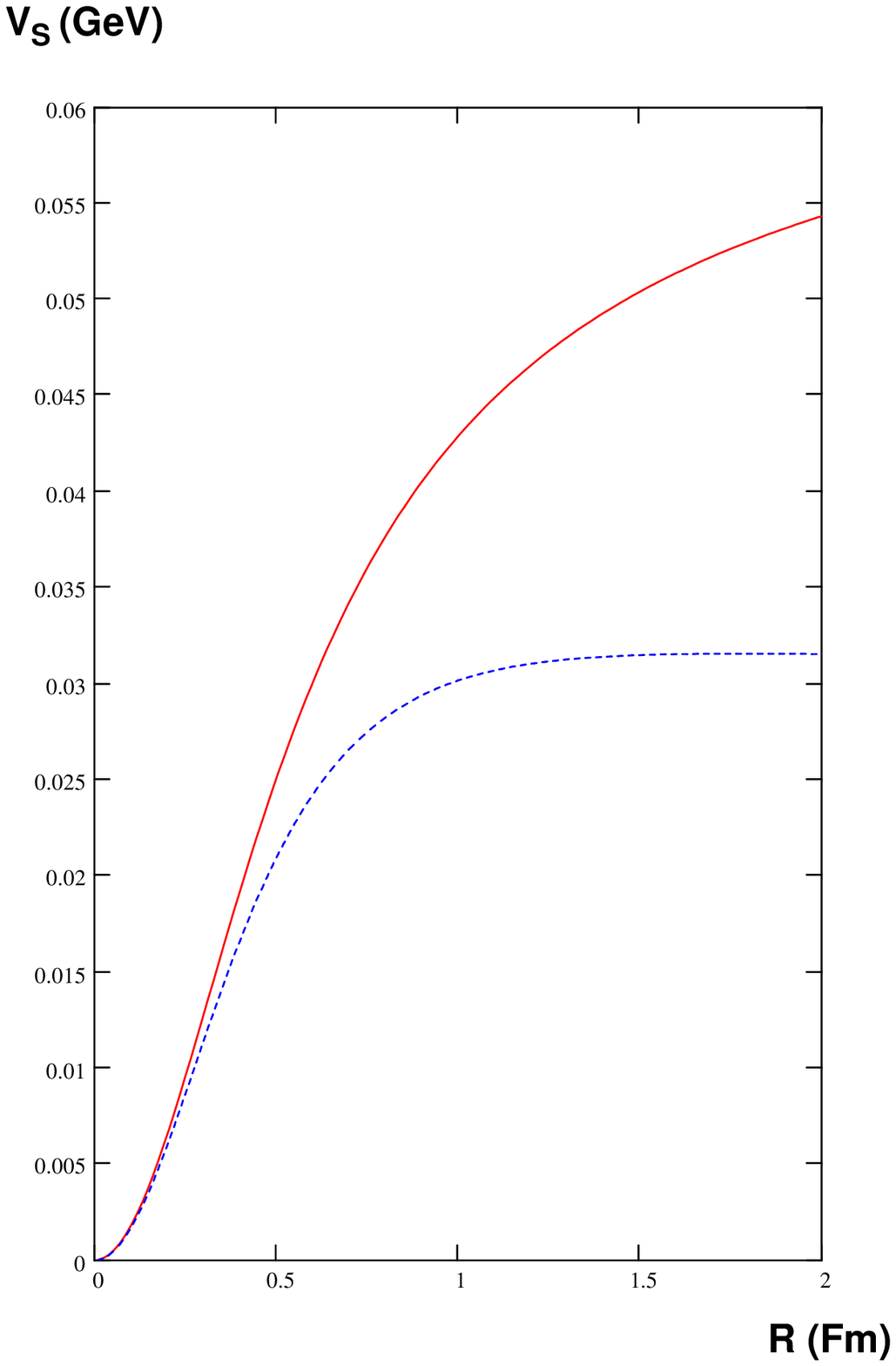}}
 \caption[dummy0]{
Scalar interaction potential.
The legend is the same as in Fig. 1
  \label{fig:S} }
\end{minipage}
\end{figure}
\begin{figure}[tbp]
\hspace*{1.cm}
\begin{minipage}{2.5in}
\vspace*{0.5cm}
\epsfxsize=6cm
\epsfysize=9cm
\centerline{\epsfbox{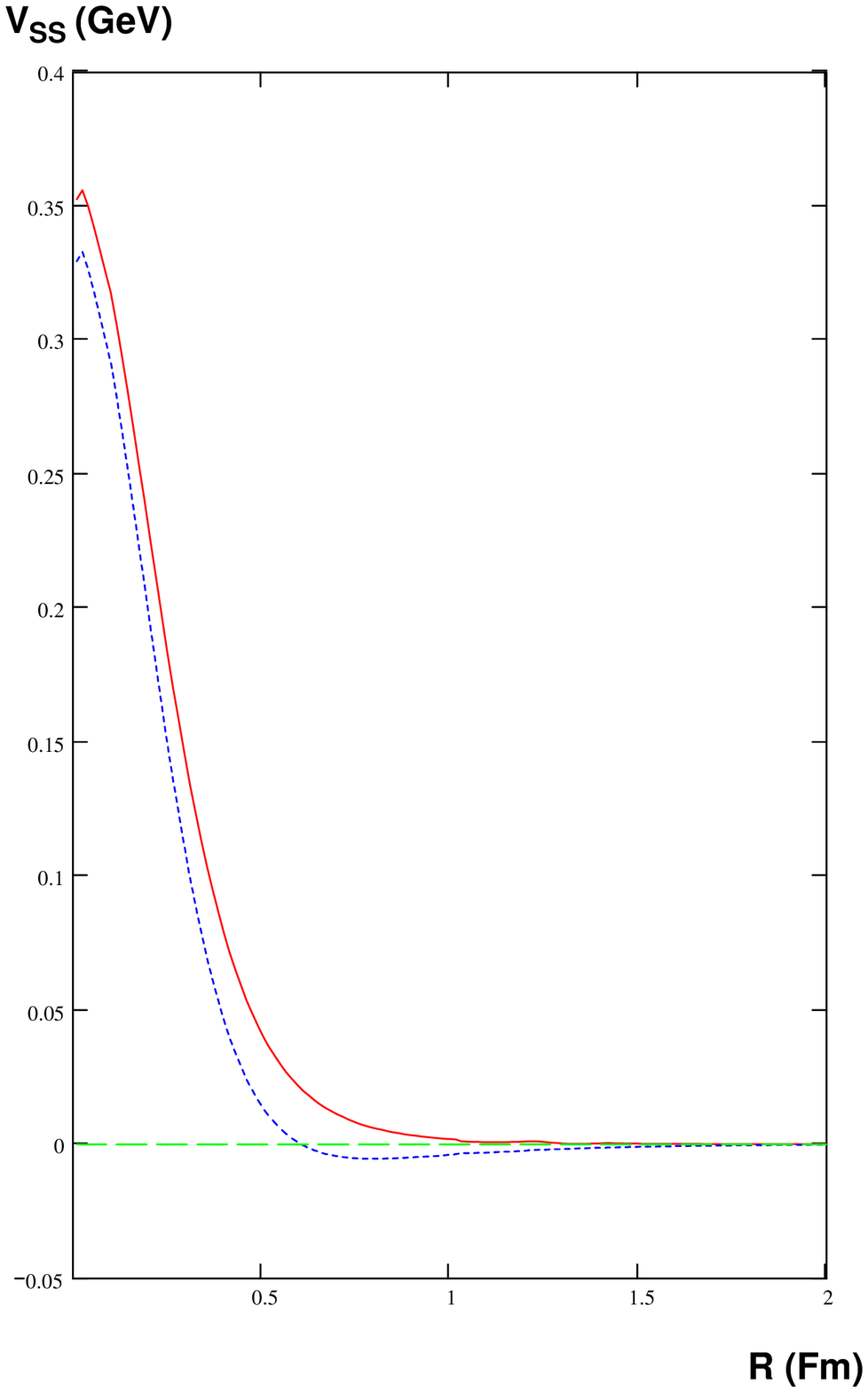}}
 \caption[dummy0]{
Spin-spin interaction potential.
The legend is the same as in Fig. 1
  \label{fig:SS} }
\end{minipage}
\hspace*{0.5cm}
\begin{minipage}{2.5in}
\vspace*{0.5cm}
\epsfxsize=6cm
\epsfysize=9cm
\centerline{\epsfbox{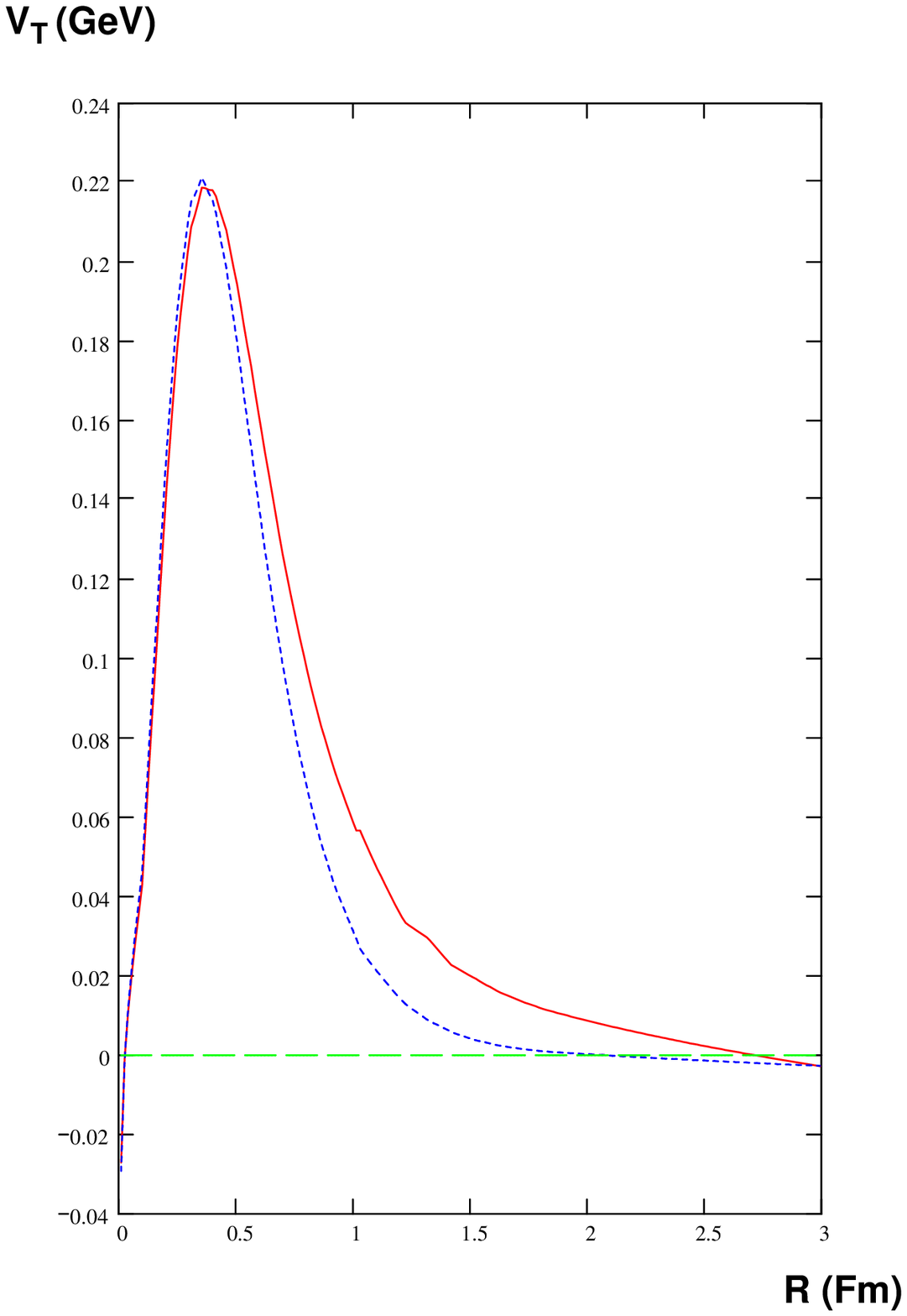}}
 \caption[dummy0]{
Tensor interaction potential.
The legend is the same as in Fig. 1
  \label{fig:T} }
\end{minipage}
\end{figure}

\section{Conclusion}

In this talk we illustrate few examples of how instanton physics works in
description of structure of QCD\ vacuum and light hadrons. In particular we
demonstrate that instantons are responsible for nonlocal properties of
vacuum condensates. Next, we show that the effective quark interaction due
to instanton exchange gives to spin-spin splittings within the hadron
multiplets. The same interaction leads to nontrivial spin-flavor structure
of nucleon sea providing large violation of isovector symmetry of sea and
essentially reducing the observed spin carried by charge constituents. Both
results was confirmed experimentally. In the last part we considered the
instanton effect on the potential between two infinitely heavy quarks. The
instanton interaction gives only renormalization of heavy quark mass, but
not to confinement of quarks. At the same time it provides small spin
effects on heavy quark potential. All this show that the instanton physics
is rather important part of nonperturbative aspects of QCD at intermediate
energies and need further considerations.

\section{Acknowledgements}

We are thankful the organizing committee for invitation to make a talk and
very nice condition for discussions. Research supported in part by the
Russian Foundation of Basic Research grant 01-02-16431 and INTAS-00-00366.

\end{document}